# Theoretical Study of the Structural, Electronic, Mechanical, and Optical of Transition Metal (Mn, Co, and Ni) Doped FrGeI$_3$ Perovskites

Nazmul Hasan[1], Alamgir Kabir[2][φ]
[1]Department of Electrical and Computer Engineering, North South University, Dhaka-1229, Bangladesh, nzharis97@gmail.com
[2]Department of Physics, University of Dhaka, Dhaka-1000, Bangladesh, alamgir.kabir@du.ac.bd
[φ] Corresponding author: alamgir.kabir@du.ac.bd

**Abstract:**

Emergence of inorganic metal halide perovskites as multifunctional optoelectronic materials are due to their exceptional tunability in optoelectronic properties. This study sought to enhance the physical and mechanical properties of lead-free FrGeI$_3$ perovskites by introducing transition metal dopants (Mn, Co, and Ni). First-principle calculations based density functional theory (DFT) have been utilized to illustrate the impact of transition-metal doping on the structural, electronic, and light-matter interaction properties of FrGeI$_3$. The study found that transition metal doping in FrGeI$_3$ perovskites leads to an increase in the electronic bandgap leading to semiconducting behavior after phase stability confirmation. Enhanced optical and mechanical properties suggest wide industrial applications in optoelectronics to biomedical area. This study provides a sound understanding of the underlying mechanisms of transition metal doping in Fr-contained halide perovskites, which could pave the way to the headway of new optoelectronic and biomedical devices based on these materials.

## Introduction:

Metal-halide perovskites have garnered substantial interest and undergone extensive investigation in recent years due to their appealing tunable electronic, optical, and elastic characteristics, rendering them as potential contenders for utilization in diverse industrial applications such as solar cells, light-emitting diodes, laser, optical data storage, medical imaging, photodetectors, spin electronics, and so on [1]–[8]. Among the different types of perovskite materials, cubic ABX$_3$ perovskites have shown great promise as a new class of photovoltaic and optoelectronic materials for their tunable direct bandgap semiconducting nature, exceptional optoelectronic properties, high absorption coefficients, long carrier lifetimes, high quantum efficiencies and high mobility of charge carriers [9], [10]. However, the main advantage came forward with these materials is that their physical properties can be further tuned for desired applications by introducing dopants in different sites. The perovskite family comprises of numerous members due to the versatile possibility of ion substitution. The light-matter interactions of these perovskites have been the subject of numerous studies, which have aided in fundamental comprehension of their light-harvesting mechanisms and have also facilitated the development of novel strategies to enhance their efficacy for device applications. Preliminarily, investigations with perovskite materials were in a box of solar cell efficiency optimization problems coming forth from Si-based technology [11]. Zhang et al. [12] have emphasized the significant adaptability of metal halide perovskites through the amalgamation of crystal orientation, surface, and quality control techniques. These approaches can potentially be applied to a range of energy-related fields, including perovskite-based energy storage, batteries, and supercapacitors, as cost-effective and proficient storage alternatives. In recent years, applications of perovskite materials in biomedical imaging significantly increasing due to their wide absorption spectrum, low trap density, fast photoresponse, and high photoluminescence quantum yields [13], [14].

Modulation of inorganic framework [$BX_6$] principally rises the formation, dynamics and recombination of polarons rather than the dipole nature of A cations in halide cubic perovskites in which distortion of the inorganic framework directly leads to the polaron effect to the nonlinear absorption [15]–[18]. By utilizing the nonlinear optical response of perovskite materials with low excitation photon energy, Two-photon Absorption (TPA) induced emission offers significant benefits over conventional luminescence by minimizing the risk of phototoxicity in biological structures [19]. As a result of these advantages, the potential for TPA in perovskite materials is vast and can be utilized in ultrafast lasers, optical data storage, and high-resolution medical imaging [20]. Very recently, one study [21] attributes that band-edge properties of perovskite materials can have a significant influence on the light emission process leading to high-performance LEDs, and laser applications with high color purity and more than 20% of external quantum efficiency. Moreover, many researchers found cubic perovskite materials can significantly enhance the performance of transistor applications [22]–[25]. With evolution of materials research, researchers are also finding optimal perovskite compounds creating variations in structural dimensions such as, double perovskites [2], [26]–[29], hybrid perovskites [30], [31], Triple-Cation Mixed/Triple-Halide Perovskite [32]–[34], two-dimensional perovskites [35]–[37] to explore new dimensions of perovskite family materials towards the technological applications by modulation materials' performance.

Despite significant efforts and advancements in the field, the instability of perovskite crystals and their potential toxicity persist as major hindrances to their further development and increased efficacy in device applications. Generally, perovskites are sensitive to temperature changes that can lead to structural phase transitions. In view of that, a tolerance factor (t) determining the perovskite crystal stability to ambient conditions has been proposed as $t=(R_A+R_X)/\sqrt{2}(R_B+R_X)$, where $R_A$, $R_B$, and $R_X$ are the A, B, and X sites' ionic radii respectively and defined that stable cubic perovskites have t values in the range of 0.8–1 [38]. The stability of materials is significantly influenced by the octahedra effect that arises between the atoms at the B and X sites. Considering this fact, larger atoms typically occupy the A site to satisfy the tolerance criterion where X atoms size needs to be larger than B site atoms. Moreover, researchers are also looking for toxicity-free perovskite materials for environment friendly technology [39]. Based on these basic concerns and utilizing the advantage of tuning properties by substitution of atoms, numerous variations are unfolded till this time on perovskite materials. The properties of $ABX_3$ structured perovskites have been the subject of numerous studies examining the effects of transition metal doping. For instance, in [40] authors attempted to broaden the absorption spectra to utilize the maximum solar radiation spectrum by Ni and Mn doping in B site with Ge in cubic perovskites and found that doping with Ni and Mn led to an increase in optical absorption for both phases, while another study [41] observed that doping with Ni led to a significant enhancement in the optical conductivity at lower energy for both lead based and lead free halide perovskites. Various studies [42]–[45] were attempted using different mechanisms to tune the physical and mechanical stability of the perovskite materials by the community and they found significant enhancements in the desired properties focusing on industry applications.

Concerning with radioactivity of Fr, no study was done on Fr based cubic halide perovskites until our previous study [46], where we demonstrated how Fr based inorganic halide perovskites interact with light, which opens a new window to think about its applications to biomedical field such as nuclear medicine and X-ray imaging technology. Among the six studied perovskites we have found that $FrGeI_3$ has high optical absorption in both the UV and visible regions, and it has high optical conductivity [46]. To further improve the optical properties and mechanical stability of $FrGeI_3$, we have introduced transition metal doping in this perovskite and the objective of this study is to examine the behavior of $FrGeI_3$ perovskite materials when subjected to transition metal doping. We have employed density functional theory (DFT) calculations to extract the enhanced behavior by transition metal doping on the structural, electronic, and light-matter interaction properties of $FrGeI_3$ perovskites. We expect that this study will provide a deeper understanding of the underlying mechanisms of transition metal doping in $FrGeI_3$ perovskites and will pave the way for the development of new optoelectronic/medical imaging devices based on these materials along with meeting the gap of perovskite explorations with heavy metals.

## Computational methodology:

The unfloded physical properties of FrBB'I$_3$ (where B=Ge and B'=Ni, Co and Mn ) have been investigated using Density Functional Theory (DFT) calculations based on plane-wave pseudopotentials. The *Cambridge Serial Total Energy Package* (CASTEP) code of Materials Studio-7.0 was utilized in this study to perform the investigations, and the structural relaxation was validated via the *Vienna ab initio Simulation Package* (VASP) [47], [48]. The unit cells of pure and doped perovskite were constructed in the cubic form and are depicted in Fig. 1. The doped FrGeI$_3$ cubic perovskites were structured using a $2 \times 2 \times 2$ supercell of pure FrGeI$_3$, resulting in 40 atoms in translational symmetry. The geometry optimizations used the PBE functional for exchange-correlation interactions with the GGA method, and Vanderbilt's Ultrasoft pseudopotential with relativistic treatment for electron-ion interactions [49], [50]. The optimized crystal phase with minimizing electronic energy and stresses was secured using the Broyden-Fletcher-Goldfarb-Shanno (BFGS) algorithm [51]. In the pure FrGeI$_3$ and doped FrGeI$_3$ systems, electron wave functions were employed in the calculations, with a cutoff energy of 450 eV and 350 eV for the plane wave, respectively. A Monkhorst-Pack's K-point mesh was used for better convergence ($10 \times 10 \times 10$ for undoped and $3 \times 3 \times 3$ for doped crystals). The finite strain theory within the CASTEP code was employed to calculate the elastic stiffness constants ($C_{ij}$). The Voigt-Reuss-Hill (VRH) averaging scheme was used to calculate the polycrystalline mechanical parameters, along with the relevant equations, in which a maximum strain amplitude of 0.003 was set [52]. The optical properties were calculated and analyzed using CASTEP-based DFT Kohn-Sham orbitals, and relevant formulas from the literature [51]. The optimization threshold was set up in the CASTEP for unit cell structures and atomic relaxation with a maximum displacement of $5 \times 10^{-4}$ Å and other specific parameters of total energy $5 \times 10^{-6}$ eV/atom, maximum force 0.01 eV per Å, and maximum stress 0.02 GPa.

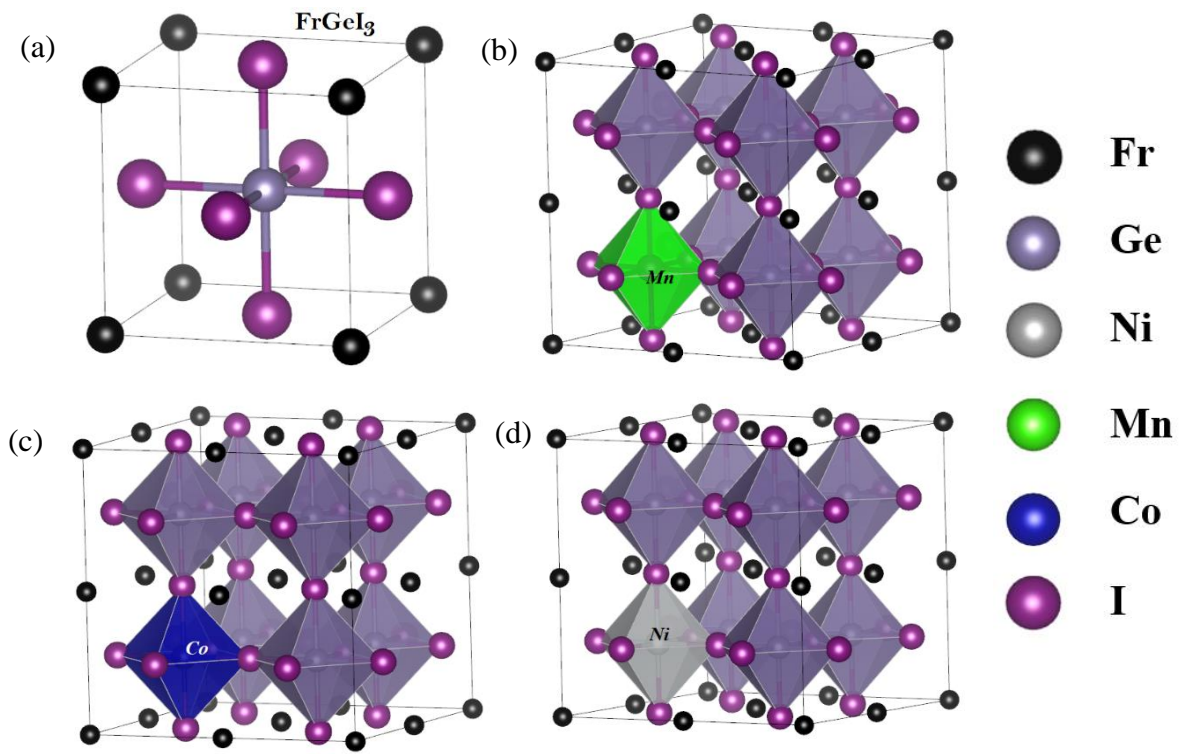

Fig. 1: DFT optimized crystal structure of the studied Fr containing perovskites (a) FrGeI$_3$, (b) FrGe$_{1-x}$Mn$_x$I$_3$, (c) FrGe$_{1-x}$Co$_x$I$_3$ and (d) FrGe$_{1-x}$Ni$_x$I$_3$. (x=12.5%)

# Physical Properties of FrGeI$_3$ Perovskites:

## A. Structural properties:

The inorganic metal-halide perovskites investigated in this study contain a highly radioactive heavy element, Fr, and crystallize in the cubic Pm$\bar{3}$m (221) phase. The unit cell of FrGeI$_3$ crystal contains one atom of Fr in A site corners at (0 0 0), one of Ge in B site center at (0.5 0.5 0.5), and three I atoms in X site face-centered positions at (0 0.5 0.5). The considered crystal structures for this study are represented in Fig. 1 which were modeled using VESTA (Visualisation for Electronic Structural Analysis) software [53]. Table 1 presents the optimized crystal's structural parameters as obtained through the calculations. The estimated lattice parameter and volume of the doped unit cell are well-tuned with the pure one which indicates their transformation periodicity in the lattice. Substitution in B site with transition metal ion results an increase in lattice parameter than pristine one while the A-X bond lengths are reduced. The atom variation with ionic radius comes up with a clear change in interatomic distances and atomic bond angles, which in turn leads to an effect on lattice constant values. It can be depicted from Table 1 that, the inorganic framework in the lattices [BX$_6$] is varied slightly by the substitution of Mn, Co, and Ni with Ge in the B site of the crystal. Variations in the Mn/Co/Ni-I bond distances lead to changes in the whole crystal structural dynamics. Decreasing nature of ionic radius for transition metal from Mn to Ni leads to a decrease in lattice parameters and volumes as presented in Table 1. The radius of B site atoms for the mixed cation is calculated using the effective mixed cation radius for metal halide perovskites:

$r_{B,Avg} = xR_{Ni/Co/Mn} + (1-x)R_{Ge}$ [54].

Following the tolerance factor criterion [54] for crystal stability, all studied perovskites exhibit cubic phase stability by forming tightly packed structures in nature signifies the small octahedral tilting.

**Table 1.** Optimized structural parameters, and evaluated tolerance factors for pristine and transition metal doped FrGeI$_3$ perovskits.

|  | FrGeI$_3$ [46] | FrGe$_{0.875}$Mn$_{0.125}$I$_3$ | FrGe$_{0.875}$Co$_{0.125}$I$_3$ | FrGe$_{0.875}$Ni$_{0.125}$I$_3$ |
|---|---|---|---|---|
| *Space group* | Pm$\bar{3}$m, No. 221 | | | |
| $a_0$ (Å) | 6.01 | 12.03 | 12.01 | 12.00 |
| $V_0$ (Å$^3$) | 217.45 | 1740.851 | 1731.934 | 1728.7346 |
| Interatomic distance (Å) | Fr–Ge: 5.21<br>Ge-I: 3.01<br>Fr–Fr: 6.01<br>I–I: 4.25 | Fr-I: 4.30<br>Ge-I: 3.01<br>Mn-I: 2.95<br>I-I: 4.25 | Fr-I: 4.28<br>Ge-I: 3.01<br>Co-I: 2.89<br>I-I: 4.25 | Fr-I: 4.24<br>Ge-I: 3.00<br>Ni-I: 2.87<br>I-I: 4.25 |
| Atomic bond angles (°) | Fr–Ge–I: 54.74<br>I–Ge–I: 90 | I-Fr-I: 59.83<br>I–Ge/Mn–I: 90 | I-Fr-I: 59.52<br>I–Ge/Co–I: 90 | I-Fr-I: 59.76<br>I–Ge/Ni–I: 90 |
| $r_A$, $r_B$, $r_X$ | $r_A$=1.80<br>$r_B$=0.73<br>$r_X$=2.20 | $r_A$=1.80<br>$r_{B,Avg}$=0.74<br>(Mn: 0.83)<br>$r_X$=2.20 | $r_A$=1.80<br>$r_{B,Avg}$= 0.73<br>(Co: 0.74)<br>$r_X$=2.20 | $r_A$=1.80<br>$r_{B,Avg}$= 0.72<br>(Ni: 0.69)<br>$r_X$=2.20 |
| Tolerance factor, T | 0.97 | 0.96 | 0.97 | 0.97 |

## B. Electronic properties:

To model the interaction of a photon with a charge carrier in crystals, it is crucial to know the electronic configuration of the material. To explain the optical behavior, it needs first to find out the bandgap energy of materials since optical properties of matter typically arise from the transition of valence electrons from the valence band to the conduction band. To explain the electronic properties of the considered intrinsic FrGeI$_3$ perovskite and how the transition metals' dopant affects their nature, we have calculated the perovskites' electronic band configurations. Orbital projected density of states also has been studied to see how an element dominates in the crystals' electronic behavior.

*Electronic Band Structure:*

The calculated band structures as illustrated in Fig. 2 (a) – (d) are evaluated utilizing the derived compounds of FrGeI$_3$ phase as illustrated in Fig. 1. The computed bandgap values are of 0.64 eV, 0.85 eV, 0.86 eV, and 0.92 eV for FrGeI$_3$ [46], Fr(Ge$_{0.875}$Mn$_{0.125}$)I$_3$, Fr(Ge$_{0.875}$Co$_{0.125}$)I$_3$, and Fr(Ge$_{0.875}$Ni$_{0.125}$)I$_3$ respectively, at the gamma point (G) of the Brillouin zone. It is clear from the Fig. 2 that the doping with transition metal Mn, Co, and Ni in B site of FrGeI$_3$ perovskites demonstrates an increasing trend in bandgap values, and that the doped band structures exhibit a shift of direct band gaps in high symmetry point from R to G and no impurity/defect states appeared in the gap region. The exhibited direct bandgap nature is necessary to be a potential candidate for photothermal-photovoltaic and optoelectronic applications as efficient photo-conversion semiconductor materials.

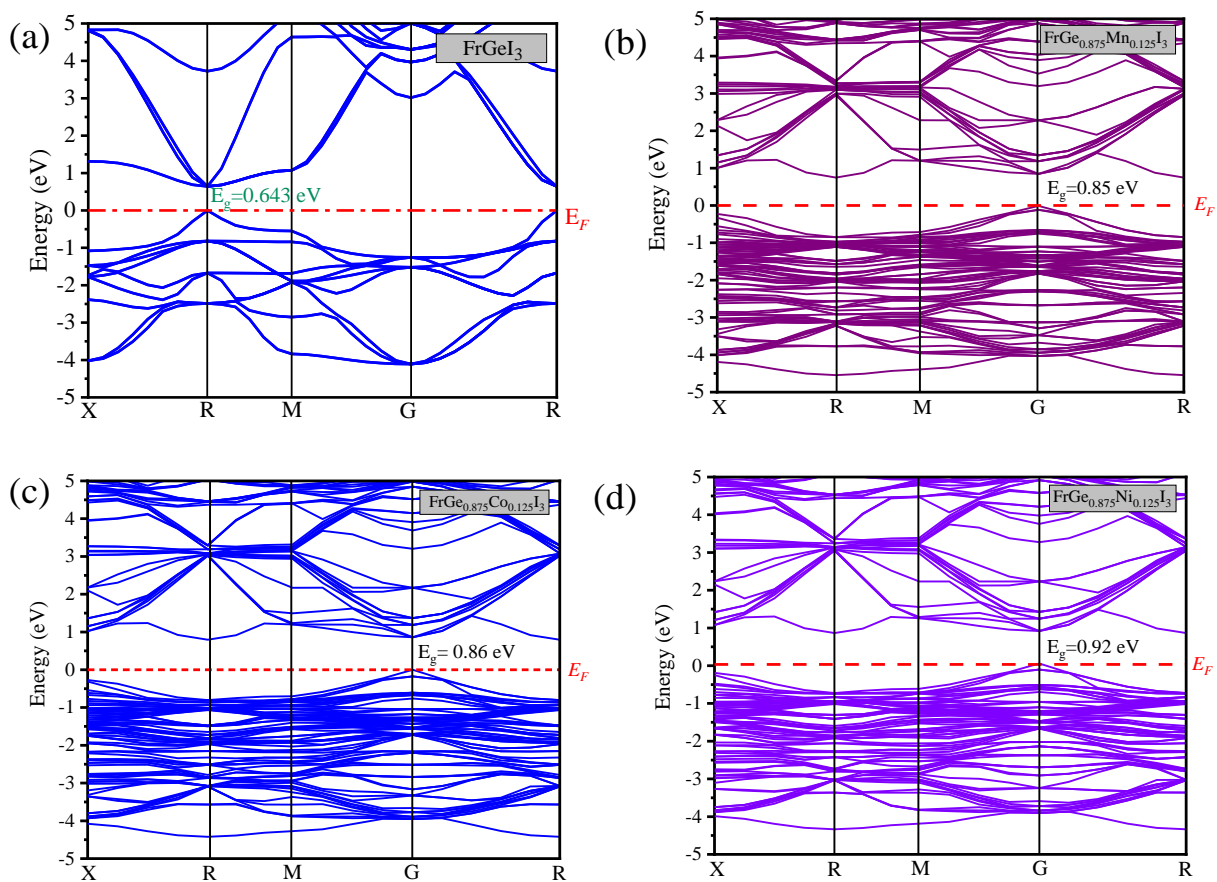

Fig. 2: Calculated electronic band structure for (a) FrGeI$_3$, (b) Fr(Ge$_{0.875}$Mn$_{0.125}$)I$_3$, (c) Fr(Ge$_{0.875}$Co$_{0.125}$)I$_3$, and (d) Fr(Ge$_{0.875}$Ni$_{0.125}$)I$_3$ cubic halide perovskites.

From Fig. 2, it is evident that substitution with transition metals Mn, Co, and Ni in place of Ge leads to higher electronic bandgaps over the pristine one and the conduction band minima (CBM) uplifts to higher energy with valence band maxima (VBM) remaining at the same energy level. With decreasing the atomic radius of the dopant atom, bandgap values increase, which can be attributed to the influence of the interatomic distance on the binding strength of the valence electrons. A lower interatomic distance may strengthen the binding forces of the valence electrons leading to a large energy requirement to excite these electrons into the conduction band, resulting in an increase in the energy bandgap as observed in Fig. 2 (b)-(d). However, other factors such as the lattice mismatch and the periodic potential field of the material must also be considered when evaluating the energy bandgap [55]. Furthermore, the dielectric constant of the material varies inversely with the energy bandgap and directly with the interatomic distance. As a result, a reduction in interatomic distance results in a rise in the energy bandgap.

Moreover, some factors that can affect in turn the electronic properties in a crystal are electronegativity, electron affinity, metallicity, and ionization energy. The substitution of atoms with lower to higher ionic radius (Mn to Ni) in B sites increases the electronegativity of the material. Increasing the electronegativity of atoms is suitable for forming chemical bonds in a system to significantly increase the ability of chemical repercussions with the interaction of light source. The trend towards increased bandgap upon doping can be attributed to the reduction in electronegativity difference between B'-site elements, such as Mn (electronegativity value of 1.55), Co (electronegativity value of 1.88), Ni (electronegativity value of 1.91), and the Iodide ion (electronegativity value of 2.96). The decrease in the electronegativity difference results in the strengthening of the covalent bond between the X-site and B-site elements. This leads to an upward shift of the B-X bonding orbital resulting in the elevation of the valence band maximum (VBM) energy. Moving from Mn to Ni, electron affinity increases as elements are in the same row of the periodic table which might affect the energy values of accepting or removing an electron from the outer shell. The direct band gap can ease the excitation of photoelectrons and career generations which is suitable for device applications. The GGA functional with local density approximation (LDA) tends to underestimate the band gap values, and researchers now frequently use alternative techniques such as GGA+U, GGA-mBJ+SOC, and HSE06 hybrid functional. Our study's objective is to explore the impact of dopings in the B site on the electronic band structure profiles, rather than to determine the precise bandgap value.

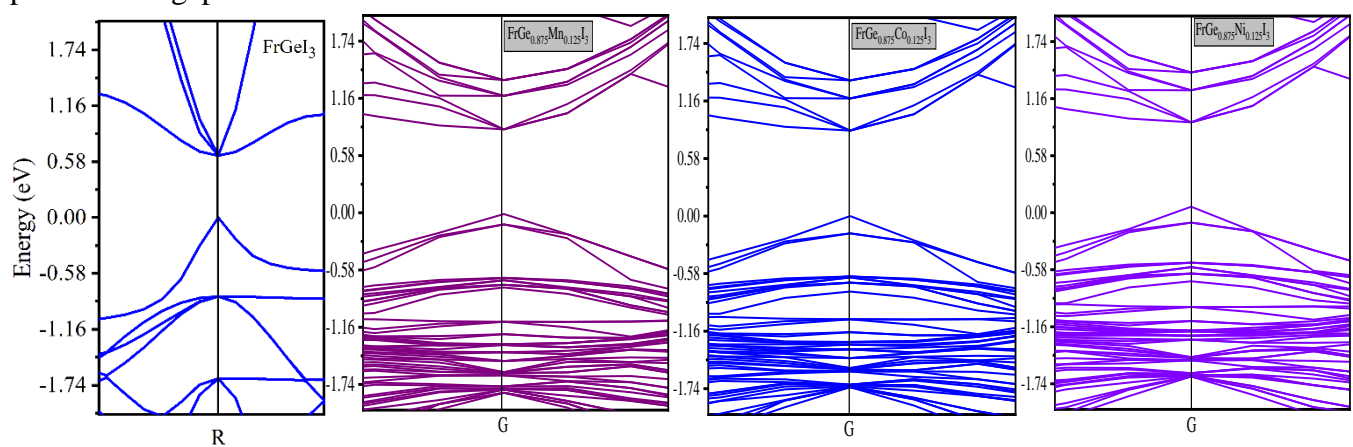

Fig. 3: Band-edge profile for the studied undoped and doped $FrGeI_3$ cubic halide perovskites.

The interaction between atomic orbitals constructs the band-edge structure of metal-halide perovskites (MHPs) which is crucial in controlling the optoelectronic characteristics of materials. To realize the band-edge influence on optoelectronic properties of the pristine and doped $FrGeI_3$ perovskites, it needs to understand correlations among band-edge electronic structures, Urbach trail, rashba splitting, and exciton binding energy [21]. Considering this importance, band-edge profiles for the studied perovskites are represented in Fig. 3. Since fundamentally optoelectronic properties originate from the band-edge orbital hybridization for MHPs, it requires to analyze the density of states to understand the band-edge profiles. Atom and orbital projected partial electron density of states from Fig. 4 shows that, valence bands originate primarily from the s-orbitals of Ge and the p-orbitals of I, whereas the conduction band is mainly associated with the p-orbitals of the B site atoms. Urbach tail and Rashba splitting can describe the thermal and structural defect resulting in absorption edge broadening and electronic structure splitting near band-edges which can significantly impact on light emission process to be utilized for light-emitting diodes, laser, and spintronics applications. However, from Fig. 3 and 4, as the energy of the electronic states approaches to the band gap energy, the electron density of states becomes increasingly localized in the band edge region which leads to a tail of electronic states with energies just below the band edge. The existence of localized states within the band gap of all perovskite materials causes the absorption spectrum near the band edge to widen due to the Urbach tail effect [56]. The Rashba effect arises due to the interplay between electron spin and momentum under the influence of an external electric field that is noticeable in materials with spin-orbit coupling and an absence of inversion symmetry, leading to the splitting of the conduction band valleys to momentum space emerging device applications of perovskites [21], [57]. For all the FrGeI$_3$ perovskites from Fig. 3, it is observed that spin-degeneracy states appeared in the conduction band for all perovskite phases indicates that the material has a weak spin-orbit coupling. In such cases, excitonic binding energies influence the studied perovskites' optical properties.

*Electron Density of States:*

The electronic density of states (DOS) provides information about the material's atomic behaviors which are instrumental in determining its physical properties, such as bandgap, conductivity, and optical properties [58]–[60]. The total-density of states (TDOS) provides insight into the energy distribution of electrons and holes within the material, which helps in understanding the charge transport properties and the material's potential for optoelectronic applications [61]. The absorption spectra of a material should align with the available energy states in its density of states (DOS) that can also provide insight into how impurities, defects, and chemical doping influence the electronic structure of the material [62]. Partial density of states (PDOS) is a refined form of TDOS that gives information about the contribution of specific atomic orbitals to the overall electronic structure which is essential in understanding the contribution of electrons in the material's opto-electronic properties. The PDOS identifies the electronic states responsible for energy dependent properties of photon for controlling photon-matter interactions at the quantum level. Insight into the bonding nature among constituent atoms reviling any hybridization effects on properties can be gained through PDOS, by determining which orbitals contribute to valence and conduction bands [63].

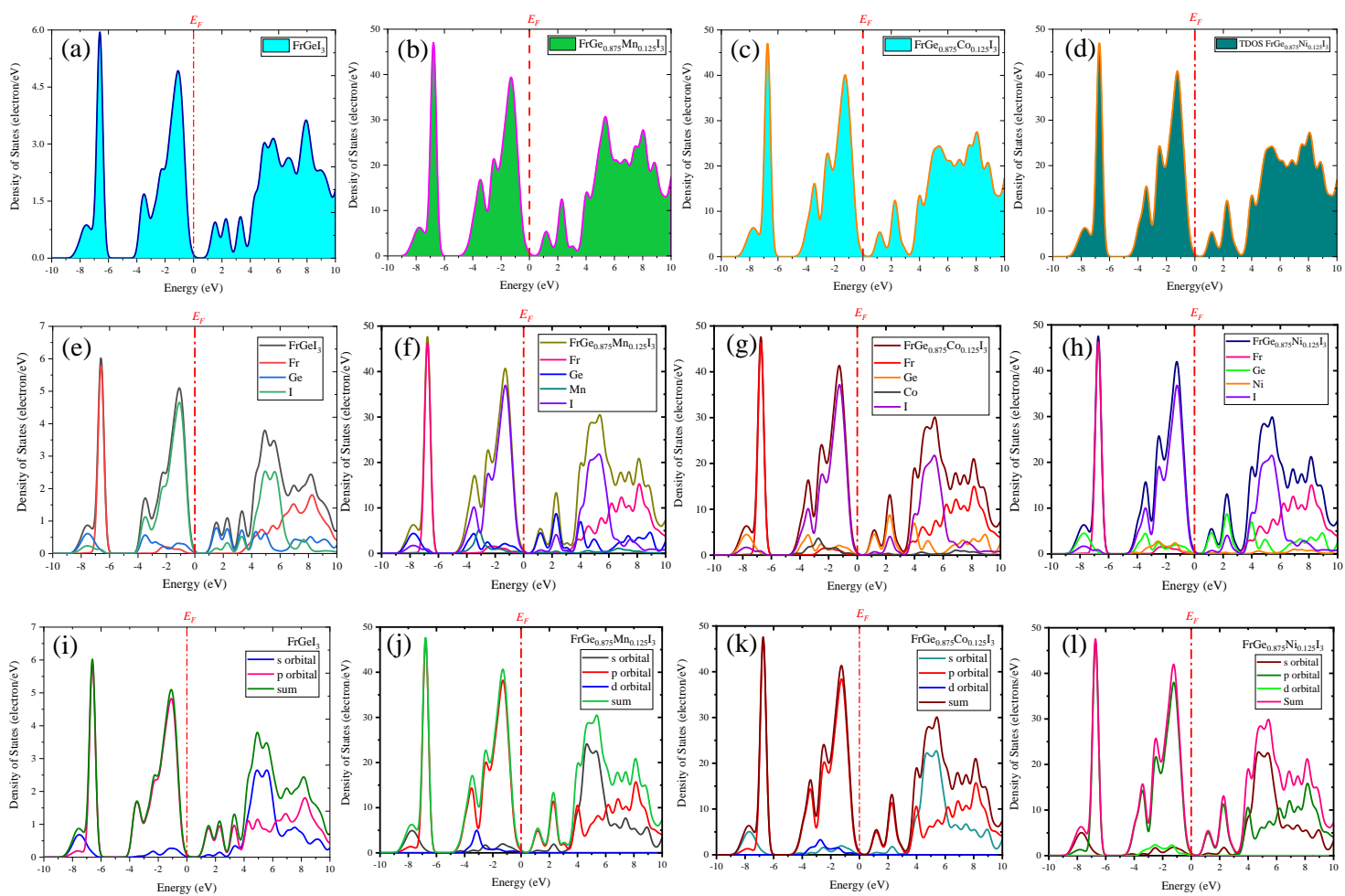

Fig. 4: Electronic properties: calculated total density of states (TDOS) plot (a)-(d), partial density of states (PDOS): atom projected (PDOS) (e)-(h) and orbital projected PDOS (i)-(l) for pristine and Mn, Co, Ni doped FrGeI$_3$ perovskites phases.

Fig. 4 displays the TDOS and atom-orbital projected PDOS of the materials and their constituent elements where it can be depicted that the valence band electrons dominate more towards the Fermi energy region than the conduction bands. The halogen atoms have a greater effect on the top of the valence band, whereas the B-site metal atoms have a greater impact on the bottom of the conduction band. The p-orbital of the I atom in the halogen site contributes significantly to the Fermi level, as depicted in the atom and orbital projected density of states (Fig. 4(e) - 4(l)). In pure $FrGeI_3$ crystal, the valence band maxima is primarily made up of the s-orbital of the Ge atom. However, doping of Mn, Co, and Ni results in the mixing of negligibly d-orbital in the valence band edge. A notable thing is showed up for the studied perovskite is that the contribution of Fr in the near Fermi energy region is zero for both VB and CB. Below the Fermi energy level, the primary contribution was from I's p orbital, while above the Fermi energy level, the s and p orbitals of B site atoms were the main contributors. A significant contribution of 5p orbital of I atom, lower electronic states of Ge atom, 4s (for all)/ 3d states of Mn, Co, and Ni atom appears to the top region of the VB (0 to -5 eV region), whereas the Ge 4p and 4s for Mn, Co, Ni states are the notable contributors to the bottom region of the CB (0 to 5eV region). Also, from any distortion of the electron cloud, non-linear optical properties may originate which can basically be determined by conduction band energies that are contributed by mixed s and p orbital of the studied $FrGeI_3$ perovskites [20]. Figures 4(f)-(h) and 4(g)-(l) demonstrate that the introduction of Mn, Co, and Ni in B-site atoms leads to a rise in electronic states below the Fermi level in the valence band due to the contribution of d orbital electrons. In addition, Fr atoms make a significant contribution to both the deep valence band and conduction band energies at -6 to -8 eV and 5 to 10 eV, respectively. The electronic configuration-based properties of the materials at the Fermi energy are not influenced by any hybridization. With doping by Mn, Co, and Ni in the B site Ge atom, the contribution of the atomic orbital near the Fermi region has changed as observed from Fig 4 (i)–(l). Concerning Mn doping, the atom-orbital projected partial density of states shows that, an additional peak appeared with a higher magnitude in the valence band than only Ge contained perovskite while in case of Co and Ni doping, the peak appeared to be sharper with higher magnitude. However, another distinct change is observed in the conduction band's lower energy region (0-3 eV) by rising to identical sharp peaks by doping in materials with higher electron densities for all cases. Substitution from Mn to Ni results in a relatively smaller decrease in electron density in the higher energy region of the conduction band.

## C. Elastic Constants and Mechanical properties:

To realize a material's mechanical properties, including its ability to bind atomic planes together, stiffness-hardness, and stability, it is essential to have a comprehensive understanding of its elastic constants. These parameters are essential in determining how a material will behave when subjected to mechanical stress, and therefore an important factor for industrial applications.

Three independent elastic moduli for the considered transition metal doped $FrGeI_3$ cubic perovskites are evaluated utilizing the finite-strain method in CASTEP and listed in Table 2. Because the perovskites being studied possess cubic symmetry, their elastic constants must fulfill the Born stability criteria to be considered mechanically stable following: $C_{11} + 2C_{12} > 0$, $C_{44} > 0$, and $C_{11} − C_{12} > 0$ [64]. The pure and transition metal (Ni, Co, and Mn) doped $FrGeI_3$ compounds confirmed mechanical stability by satisfying the above mentioned criteria as observed in Table 2. Moreover, another essential parameter, Cauchy Pressure ($C_{12}$-$C_{44}$), is evaluated for the considered perovskites to determine whether they are brittle or ductile[44]. Pure $FrGeI_3$ shows whiley brittleness where transition metal doped compounds demonstrate ductile materials as signifies by their positive Cauchy pressure values. The same type of transition (brittleness to ductility) was observed previously on Mn/Ni/Co doped halide perovskites [41], [65], [66]. Elastic compliance ($S_{ij}$) is the ratio of strain to stress in a material that represents the degree to which a material deforms under applied stress. A material with a high elastic modulus will have a low elastic compliance and vice versa which is clearly reflected for the studied perovskites in Table 2. These values reconfirm the least and negligible material deformation under plastic deformation suits industrial applications.

To quantify the studied perovskite materials' mechanical behavior with strain-stress, necessary parameters are evaluated using the Voigt–Reuss–Hill (VRH) approximation i.e., Bulk (B) and Shear (G) Moduli, Young's Modulus (Y), Pugh's ($B/G$) and Poisson's ($v$) Ratios, Debye Temperature ($\Theta_D$), Anisotropy (A), and Compressibility (k). The bulk modulus for all of the (Mn, Co, and Ni) doped and undoped $FrGeI_3$ perovskites are found in the range of 15.4 GPa to 21.5 GPa, which are small in magnitude indicating their softness and flexibility [43]. Such values suggest that, these perovskites can be utilized in optical thin film applications. The shear modulus G measures a solid's resistance to shape deformation parallel force onto its surface, while Young's Modulus (Y) represents its stiffness and resistance to deformation by perpendicular stress to its surface. Table 2 provides evidence that the transition metals doping in the $FrGeI_3$ halide materials leads to an increase in their shear-stiffness values and indicates that the doped materials are more resistant to shape change and less likely to undergo plastic deformation. Therefore, the mechanical stability of the considered perovskites is increased by Mn, Co, and Ni doping in the B site, making them suitable for applications that require high resistance to mechanical stress. Considering both Pugh's and Poisson's Ratios, transition metal Mn, Co, and Ni doped Fr based iodide perovskites exceed the threshold values (1.75 and 0.26 respectively) [65] to be distinguished as ductile materials.

**Table. 2**: Evaluated elastic constants and mechanical parameters for transition metal doped and undpoed FrGeI3 perovskites.

| Compound | $C_{ij}$ | | $C_{12}$-$C_{44}$ | $S_{ij}$ | B | G | Y | B/G | $v$ | $\Theta_D$ | A | k |
|---|---|---|---|---|---|---|---|---|---|---|---|---|
| | i j | GPa | | 1/GPa | GPa | | | | | K | | |
| $FrGeI_3$ | 1 1 | 33.17 | -0.96 | 0.032 | 15.42 | 9.46 | 31.01 | 1.63 | 0.24 | 126.13 | 0.40 | 0.062 |
| | 1 2 | 6.55 | | -0.005 | | | | | | | | |
| | 4 4 | 7.51 | | 0.133 | | | | | | | | |
| $FrGe_{0.875}Mn_{0.125}I_3$ | 1 1 | 38.38 | 0.88 | 0.028 | 17.93 | 10.07 | 35.81 | 1.78 | 0.26 | 136.11 | 0.38 | 0.056 |
| | 1 2 | 9.68 | | -0.005 | | | | | | | | |
| | 4 4 | 8.80 | | 0.114 | | | | | | | | |
| $FrGe_{0.875}Co_{0.125}I_3$ | 1 1 | 39.04 | 1.78 | 0.034 | 18.96 | 10.61 | 36.82 | 1.79 | 0.27 | 134.36 | 0.43 | 0.054 |
| | 1 2 | 10.46 | | -0.006 | | | | | | | | |
| | 4 4 | 8.68 | | 0.115 | | | | | | | | |
| $FrGe_{0.875}Ni_{0.125}I_3$ | 1 1 | 41.60 | 2.61 | 0.027 | 21.52 | 11.46 | 36.63 | 1.87 | 0.29 | 133.28 | 0.46 | 0.046 |
| | 1 2 | 11.48 | | -0.006 | | | | | | | | |
| | 4 4 | 8.87 | | 0.123 | | | | | | | | |

The Debye temperature of perovskite materials represents the upper limit of thermal energy in a material, reflecting the average energy of atomic vibrations in the crystal structure which strongly influences essential properties for designing specific devices including material's heat capacity, thermal conductivity, and melting temperature. Table 2 shows that the Debye temperature for the pure and transition metal doped $FrGeI_3$ perovskites range above 125 K to 136 K which implies that strong atomic vibrations resulting to perfect thermal conductivity and hence the optimal device efficiency [67]. The anisotropy index relates to the directional dependence of the mechanical, electrical, and magnetic properties in perovskite materials [64], [68], which is significant in a variety of applications, such as phase transformations, and dislocation dynamics. For cubic $FrGeI_3$ perovskite crystals, the degree of anisotropy is quantified using the anisotropy factor (A) defined by a well-established formula as reported by Ranganathan and M. Ostoja-Starzewski[68], and the values (in Table 2) for pure and Ni, Co, Mn-doped compounds show that all A values are nonunity, indicating an anisotropic nature for the studied perovskite materials [66]. Additionally, the A values from Table 2 suggest that Ni-doped perovskite has higher anisotropy than the other doped phases where the Mn-doped phase demonstrates the smallest anisotropy, making it more suitable for device applications[65], [66], [69].

Compressibility (*k*) is a fundamental elastic property that describes the ability of a material to undergo compression when subjected to an external force, and it is defined as the fractional change in volume of the material in response to applied pressure. According to Table 2, the low and negligible values (0.04-0.06 range) of the compressibility (*k*) suggest that these materials have the ability to withstand high-pressure loads without experiencing permanent deformation, making them promising candidates for use in various technological applications [70]. Moreover, perovskite materials exhibit piezoelectric properties, which is related to their compressibility, as the materials' ability to undergo mechanical deformation is closely related to their ability to generate an electric field under applied stress. Furthermore, the compressibility of perovskite materials is also influenced by the direction of applied stress, which can be exploited in designing devices that utilize the piezoelectric effect [71]. Since, the compressibility of perovskite materials can be controlled through doping as observed from Table 2 values by transition metal doping in $FrGeI_3$ perovskites, which can have a significant impact on the materials' piezoelectric properties.

### D. Light-matter interactions:

The manner in which perovskite materials interact with light is crucial in determining their efficacy in optoelectronic applications such as solar cells and light-emitting devices. Perovskite materials absorb photons upon exposure to light results in generating electron-hole pairs through electron excitement, and subsequent relaxation of excited electrons governed by a complex interplay of carrier generation, transport, and recombination mechanisms. Photovoltaic devices produce a photocurrent by separating excited electrons and holes with an internal electric field and directing them to the electrodes. On the other hand, light-emitting devices emit light by allowing electrons and holes to recombine, and the emitted light can be controlled by modulating the perovskite material's bandgap [72]. Moreover, the excellent optical properties of perovskite materials, including high photoluminescence quantum yields, tunable emission wavelengths, and high brightness, can improve the efficiency and sensitivity of biomedical imaging techniques [73], [74]. The tunable optical behavior of perovskites can be used to target specific biological structures by selecting the appropriate excitation wavelength and resulting in stable x-ray imaging [75]. Also, the high brightness of perovskite materials makes them attractive candidates for photoacoustic imaging, as the signal-to-noise ratio can be improved by enabling deeper tissue penetration [76]. Hence, understanding and controlling this interaction is essential for optimizing the performance of perovskite devices and unlocking their full potential for future applications from optoelectronics to biomedicine.

Considering the importance of perovskite materials' optical properties, we have studied how B-site doping can influence the pure $FrGeI_3$ perovskite's optical properties. The B site transition metal cation choice affects perovskite materials' electronic and optical properties, depending on the cation's interaction with surrounding anions (octahedral framework) [18], [77], [78]. Transition metal cations at the B site of perovskite materials can create defect and charge transfer states that affect absorption, emission, and charge transport efficiency, with implications for electronic and optoelectronic device design [79]–[81]. Density functional theory (DFT) simulations were utilized to extract the absorption spectra (light energy and wavelength dependent), including index of refraction, dielectric constant, optical conductivity, and reflectivity. The profile of the absorption coefficient α(ω), for the considered perovskites are evaluated through: $\alpha(\omega) = \frac{\sqrt{(2\omega)}}{c}\sqrt{[(K) - \varepsilon_1(\omega)]}$, where $K = \sqrt{(\varepsilon_1^2(\omega) + \varepsilon_2^2(\omega))}$, and $\varepsilon_1$, $\varepsilon_2$ respectively the real, imaginary part of the complex dependent dielectric function. Complex frequency-dependent dielectric function ε(ω) is determined as: $\varepsilon(\omega) = \varepsilon_1(\omega) + i\varepsilon_2(\omega)$. Moreover, polarization and susceptibility for the perovskites can be determined using the real part of dielectric parameters following relation: $P(\omega) = \chi(\omega)\varepsilon_0 E(\omega) = [\varepsilon_1(\omega)-1] \varepsilon_0 E(\omega)$. Considering liner optical behavior, oscillating electric field produced polarization is also with the same frequency $\omega$ in all cases which in turn utilize low to medium light intensities of EM wave [82]. However, other necessary optical parameters are refractive index n(ω), extinction coefficient k(ω), and optical conductivity σ(ω) are evaluated using the following relations [29]:

$n(\omega) = \frac{1}{\sqrt{2}}\sqrt{[(K) + \varepsilon_1(\omega)]}$ ; $k(\omega) = \frac{1}{\sqrt{2}}\sqrt{[(K) - \varepsilon_1(\omega)]}$ ; $\sigma(\omega) = \frac{\omega \varepsilon_2}{4\pi}$ ;

The calculated optical absorption profiles for pure and transition-metal doped FrGeI$_3$ perovskites are represented in Fig. 4. All the calculated absorption profiles exhibit their identical signature from ultra-violet (UV) to near-infrared (NIR) region. Noteworthy, the absorption profile for the considered FrGeI$_3$ perovskites quite agrees with the fundamental energy bandgap as displayed in Fig. 2. Within the visible photon energy (1.5-3.1 eV) range, the absorption profile for the perovskites steadily increases with the incident photon energy. However, three major peaks are observed in the whole energy range for all compositions, with the maximum absorption peak occurring at 13 eV in the UV region which is particularly useful in applications such as photodynamic therapy, where the absorption of light in this region can activate photosensitizers and spawn reactive oxygen to destroy cancer cells [83]. As, Mn doped FrGeI$_3$ demonstrates this highest magnitude in absorption profiles at this higher energy indicates its suitability to be a good candidate for biomedical applications such as photodynamic therapy, and imaging [84]–[87].

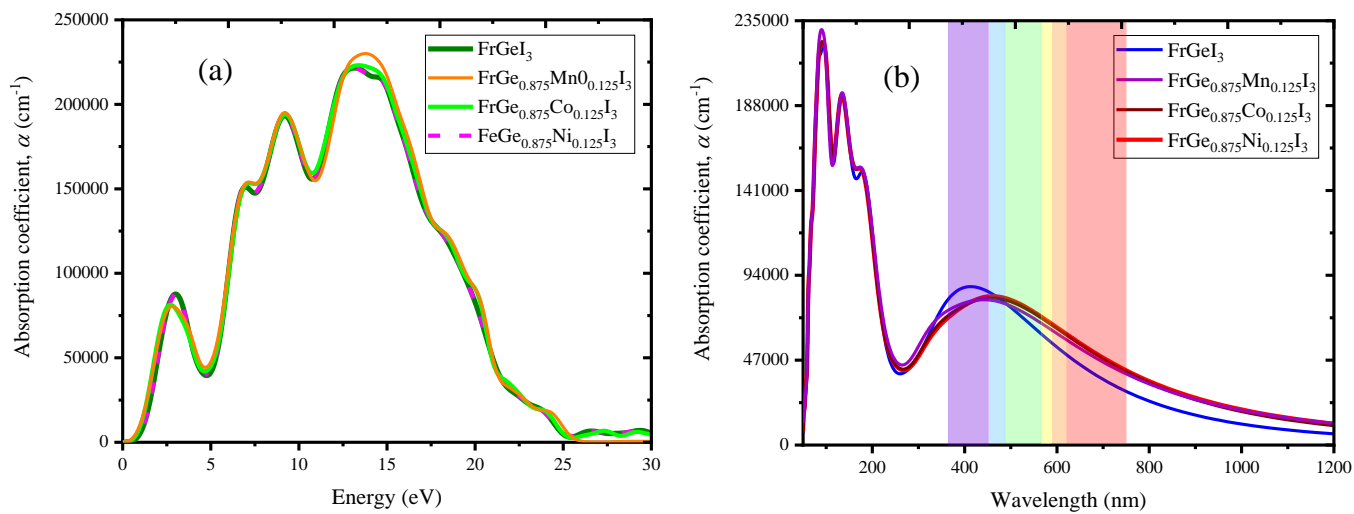

Fig. 4: DFT extracted optical absorption profile (a) as a function of energy and (b) as a function of wave length for the considered Fr(GeX)I$_3$ (X= Ni, Co, Mn) based cubic halide perovskites.

Within the low photon energies of the visible region, the undoped FrGeI$_3$ perovskite displays larger peak value than the doped one. Specifically, FrGeI$_3$ perovskite shows dominance at the transition from UV to visible spectra and is highest for absorbing violet and blue colors of visible lights (400 nm-480 nm) as can be depicted in Fig. 4(b). In contrast, the Ni and Co doped FrGeI$_3$ perovskites demonstrate higher absorption values in the region of 480 nm - 600 nm for green, yellow, and orange colors and from 600 nm to 700 nm for the red light of the visible spectrum, respectively.

As individual quanta, photons interact with individual electrons/atoms and the electrons binding energy in the band edges is strongly influenced by the inorganic framework that can shift the absorption peak towards higher or lower energies [88]. In addition, the increased binding energy of electrons by more localized electrons in conduction bands for the studied perovskites as shown in Fig. 2 results in a higher electron density and thus enhanced absorption of light. Energetic levels of the exciton can be approximated by building a relationship with binding energy and band gap values by: $E_n = E_g - \frac{E_b}{n^2}$ [89]. In direct bandgap semiconductors, the absorption spectrum at the bandgap region can be characterized by the Elliott formula that combines the influence of exciton and continuum transitions of bands related to bandgap [90]. The studied perovskite materials showed a shift towards higher energies in their absorption profile with a change in B site from Mn to Ni, indicating the effect of ionic radius changes on the structure. Additionally, Fig. 4(a) and 4(b) show that changes in optical absorption due to cation substitution in the B site that can be ascribed to the quantum-confinement effect of 3D perovskites [91]. The unique band structure and crystal symmetry of inorganic cubic iodide perovskites can result in strong excitonic effects that make them suitable for designing efficient light-emitting devices.

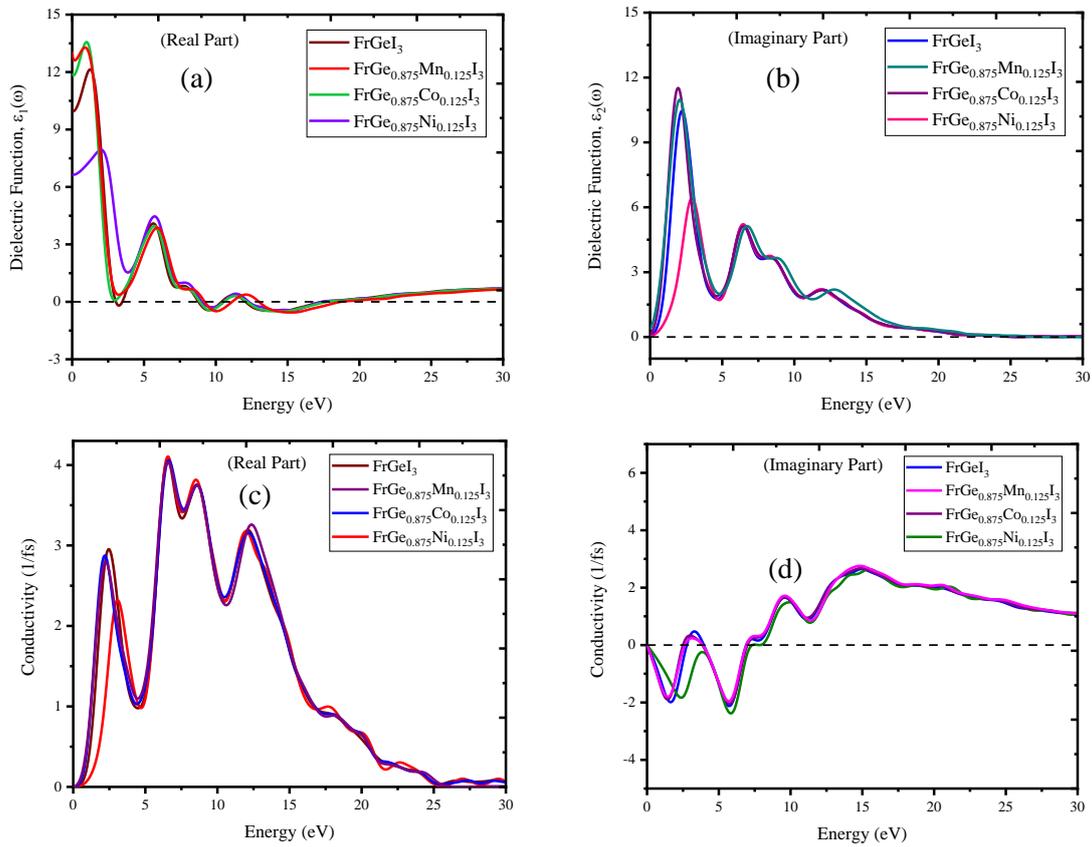

Fig. 5: Computed optical profiles for the considered Fr(GeX)I$_3$ (X= Mn, Co, Ni) based cubic halide perovskites- Dielectric Functions (a) Real part & (b) Imaginary part, and Optical conductivity (c) Real part & (d) Imaginary part.

In inorganic cubic halide perovskites, the dielectric constant varies with the crystal composition of the material. Fig. 5 (a) and 5(b) contains the real part (RCDF) and imaginary part (ICDF) of complex dielectric functions for the studied metal doped and undoped FrGeI$_3$ perovskites. The real part of dielectric constants describes the polarization of the crystals whereas the imaginary part describes the energy dissipation behavior of the material. As observed from Fig. 5(b), ICDF at E=0 eV for pure FrGeI$_3$ and Ni doped FrGeI$_3$ compounds and a negligible magnitude (~0.5 eV) was observed for Mn and Co doped compounds. Considering the negligible magnitude for Mn and Co doped FrGeI$_3$ perovskite, it portrays almost no energy dissipates within these doped and undoped FrGeI$_3$ crystal systems at zero photon energy. At the photon energy E=2.5 eV the first sharp peaks of ICDF for the compounds are observed in the visible range in Fig. 5(b) and at the same energy in visible range absorption peaks are also observed in Fig. 4(a). Hence, the observed peaks for ICDF in Fig. 5(b) support the absorption peaks as found in Fig. 4 for the studied perovskite compounds. Since, materials with higher dielectric constants have relatively lower charge-carrier recombination rate that increases the optoelectronic devices' performance efficacy, it is also important to analyze the behavior of dielectric functions varying with transition metal doping. It is evident from Fig. 5(a) and 5(b) that, Mn containing FrGeI$_3$ perovskite shows a relatively higher magnitude in both the real part and imaginary dielectric function at E=0 eV which suggests that Mn-based perovskite may be a suitable candidate for optoelectronic device applications. Electronic bandgaps are increased with the B-site dopings in heavy metal Fr based halide perovskites as observed in Fig 2, where in Fig. 5 dielectric constants are also increased for Mn and Co based perovskites and decreased for Ni based FrGeI$_3$ compound. The effect of the octahedral framework of B site atoms by its ionic radius is evident for the dielectric magnitude variations with changing from Mn to Ni. However, overall dielectric behaviors are identical for all compounds as displayed in Fig. 5 in the higher photon energy region. In high photon energies ($\geq$ 20 $eV$), the imaginary part of the complex dielectric function for all phases appears with zero magnitude whereas the real part tends to be almost unity. This behavior implies that all the perovskites exhibit transparency with a little absorption in higher energy regions as also observed in Fig. 4 looking into absorption and transparency edges.

The underline mechanism of materials' interaction with light at different frequencies is determined by the perovskites' optical conductivity. Interband transitions between the valence and conduction electrons govern the optical conductivity of cubic halide perovskites at low photon energies. As the photon energy is increased, the optical conductivity can also be influenced by higher-order processes such as electron-electron interactions and excitonic effects. The calculated optical conductivity for the considered pristine and transition metal doped $FrGeI_3$ perovskites are represented in Fig. 5(c) and 5(d). For all $FrGeI_3$ perovskites, optical conductivity occurs within a wide electromagnetic spectrum from low energy to high energy regions. Mn-containing materials exhibit the highest amplitudes in the low-energy region, while Ni-containing compounds depict the highest magnitudes across the entire spectrum at 7eV. Compared to pristine $FrGeI_3$ compound, doping with Ni, Co, and Mn in the B site results to increase in the conductivity for Co and Mn in the visible energy region where Ni doped $FrGeI_3$ shows its dominance over the higher energy (UV) region.

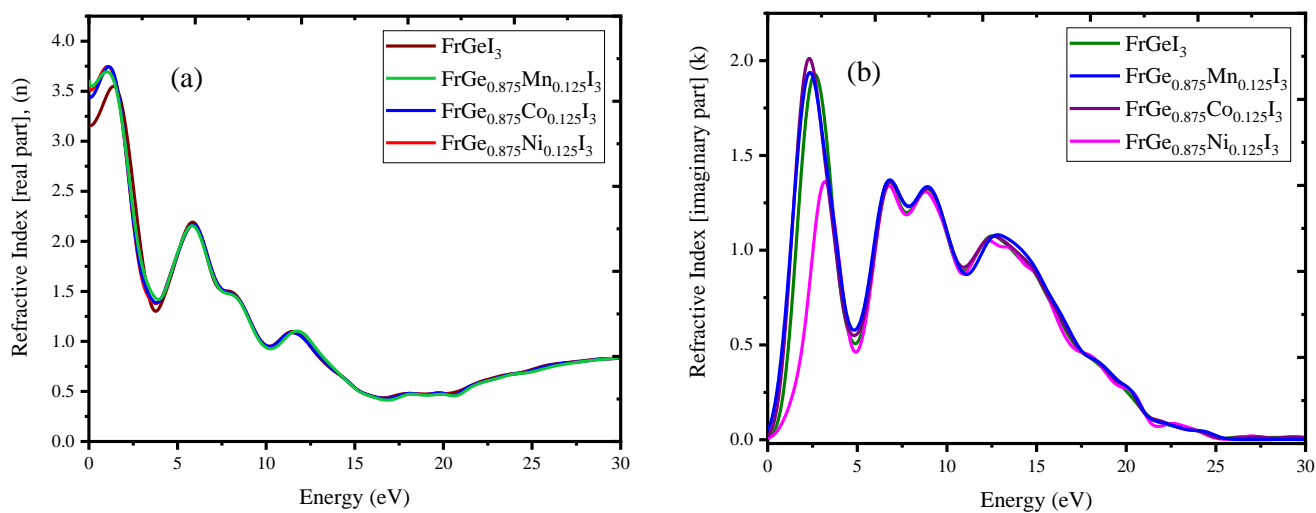

Fig. 6: DFT retrieved (a) Reflective index and (b) extinction coefficient for the considered $Fr(GeX)I_3$ (X= Ni, Co, Mn) perovskites.

The extent to which a material decelerates the speed of light as it passes through, defining the behavior of light propagation in absorbent materials, is defined by the refractive index. All of the studied perovskites have similar qualitative refractive indices (in Fig. 6), but with slight variations in peak heights and peak positions. The static refractive index ranges from a minimum of 3.15 for $FrGeI_3$ to a maximum of 3.65 for $FrGe_{0.875}Mn_{0.125}I_3$, and further transition metal doping decreases the refractive index values with changes by the atomic radius of Mn, Co, and Ni. The high refractive index (>1) in halide perovskites is due to the large electron density and narrow band gap, leading to a strong interaction between light and the material, causing light to slow down and bend as it passes through the perovskite. The polarization of the electric field component of an electromagnetic wave on a material's surface affects the dielectric functions and, in turn, the refractive indices. Besides, light speed in an absorbing medium is inversely related to the refractive index value by $\sqrt{\varepsilon_1}$. Since, electron oscillation contributes only to the dielectric polarization of material in visible light energy, it is ultimately synchronized along the refractive index values. However, imaginary part of the refractive index (k) indicates the amount of light absorption or attenuation in the material, as depicted in Fig. 6(b) that is related to the absorption coefficient ($\alpha$) following: $\alpha = \frac{4\pi k}{\lambda}$. The relation between k and $\alpha$ shows that the absorption coefficient increases as the imaginary part of the refractive index increases, indicating that materials with a higher k value will absorb more light. Fig. 5(b) and Fig. 2 demonstrate that the imaginary part of the refractive index rises as the photon energy approaches the band gap energy, indicating that more light is absorbed in these energies. Moreover, scattering occurs when photons are scattered in different directions by the material's structure which is more likely to occur at shorter wavelengths (higher photon energies), due to the smaller scattering centers comparing the wavelength of the light. Both scattering and absorption tend to increase with increasing photon energy, leading to an overall increase in the material's extinction coefficient as observed from Fig. 4 and Fig. 6(b) for the transition metal doped $FrGeI_3$ perovskites.

| Perovskite | $\varepsilon_1(0)$ | $n(0)$ | $R(0)$ | $\chi$ |
|---|---|---|---|---|
| $FrGeI_3$ | 10.00 | 3.20 | 0.274 | 9.00 |
| $Fr(Ge_{1-x}Mn_x)I_3$ | 13.00 | 3.52 | 0.324 | 12.00 |
| $Fr(Ge_{1-x}Co_x)I_3$ | 11.90 | 3.48 | 0.300 | 10.90 |
| $Fr(Ge_{1-x}Ni_x)I_3$ | 6.56 | 3.50 | 0.180 | 5.56 |

Table 2: The calculated static values for dielectric constant $\varepsilon_1(0)$, refractive index n(0), reflectivity R(0) and susceptibility ($\chi$) for the studied tansition metal doped $FrGeI_3$ halide perovskites.

Reflectivity (R) measures the amount of light reflected by a material and is dependent on factors such as the angle of incidence and polarization at the boundary between materials with differing refractive indices. The reflectivity spectra for the pristine and Ni, Co, and Mn doped $FrGeI_3$ perovskite compounds with respect to photon energy are portrayed in Fig. 7. Doping with Co and Mn in $FrGeI_3$ results in higher amplitude in the reflectivity profile where Ni doping shows opposite trend as shown in Fig. 7. In visible photon energy region, it shows the highest magnitude for Co contained Fr based cubic perovskite while in higher energy region Mn based compound dominates over all the samples. The reflectivity of a material can also be affected by its absorption properties. If the material absorbs a significant amount of light at the wavelengths of interest, then the amount of light that is reflected will be lower. Exactly this behavior can be realized in comparison between Fig. 4(a) and 6(b). Additionally, surface properties can be leading to an increase or decrease in the overall reflectivity in practical applications.

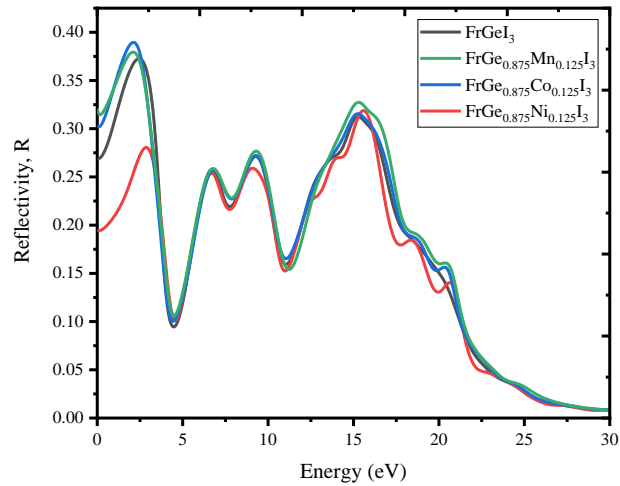

Fig. 7: Reflectivty for the considered $Fr(Ge_{1-x}T_x)I_3$ (T= Mn, Co, Ni) based cubic halide perovskites as evaluated throguh past-princple calculations.

**Summary and perspective**

Considering the growing interest in perovskites for various optoelectronics applications and biomedical technologies for their tunable optoelectronic properties by structural dynamics, we have focused to see the variations by transition metals (Mn, Co, and Ni) doping in Fr based halide perovskites. Quantum mechanical density functional theory (DFT) extracted energy bandgaps and electronic density of states show that the considered perovskites exhibit semiconducting nature with increasing trend. Doping by transition metals increases the electron density by the influence of their d-orbital leads to enhance the absorption coefficient in higher energy region while absorbance and photoconductivity in visible energy region are dominated by pristine one. Mn doped FrGeI3 perovskite shows compatible behavior in optoelectronic properties demonstrating its suitability for excellent biomedical applications. Moreover, all the studied perovskites show higher refractive index values leads to their potential for excitonic effect-based electronics such as laser, light emitting diodes, and so on. Polycrystalline mechanical behaviors remark about the ductility and strong resistivity to plastic deformation as required for industrial applications. Overall mechanical properties suggest that the materials have potential for piezo-electronics as well as flexible electronics. Based on the DFT retrieved physical properties, it is inferred that FrGe$_{1-x}$T$_x$I$_3$ (T=Mn, Co, Ni) would be potential candidates as a metal-halide perovskite for multifunctional optoelectronic and biomedical imaging applications.

**Author contributions**
N. Hasan designed and performed the calculations for this study. Both authors prepared the manuscript. A. Kabir supervised to accomplish the doping studies. Finally, all reviewed the manuscript before submission.

**Declaration**
There is no conflict of interest to declare.